\documentclass[fleqn,11pt]{article}

\usepackage[dvips]{graphicx}
\usepackage{color}
\usepackage{amsmath}

\def\NN{\hbox{I\kern-.2em\hbox{N}}}
\def\RR{\hbox{I\kern-.2em\hbox{R}}}

\definecolor{myyellow}{rgb}{0,.4,1.0}

\title{How specific exponential type orbitals recently became a viable basis
set choice in NMR shielding tensor calculation.}

\author{Philip E. Hoggan}

\begin{document}

\maketitle





\vspace{2mm}

\centerline{ LASMEA, UMR 6602 CNRS.}

 \centerline{University Blaise Pascal}

\centerline{ 24  avenue des Landais,}

\centerline{ 63177 AUBIERE Cedex, FRANCE.}


\begin{abstract}
This paper advocates use of the atomic orbitals which have direct
physical interpretation, i.e. Coulomb Sturmians and hydrogen-like
orbitals. They are exponential type orbitals (ETOs). Their radial
nodes are shown to be essential in obtaining accurate nuclear
shielding tensors for NMR work.
The present work builds on a 2003 French PhD and many numerical results were
published by 2007. The improvements in this paper are noteworthy, the key
being the actual basis function choice.

Until 2008, their products on different atoms were difficult to
manipulate for the evaluation of two-electron integrals. Coulomb
resolutions provide an excellent approximation that reduces these
integrals to a sum of one-electron overlap-like integral products
that each involve orbitals on at most two centers. Such two-center
integrals are separable in prolate spheroidal co-ordinates. They are
thus readily evaluated. Only these integrals need to be re-evaluated
to change basis functions.

In this paper, a review of the translation procedures for Slater
type orbitals (STO) and for Coulomb Sturmians follows that of the
more recent application to ETOs of a particularly convenient Coulomb
resolution.

\end{abstract}

\rule{17cm}{0.1mm}

\newpage

\section{Introduction}
The criteria for choice between gaussian and exponential basis sets
for molecules do not seem obvious at present. In fact, it appears to
be constructive to regard them as being complementary, depending on
the specific physical property required from molecular electronic
structure calculations.

The present work describes a breakthrough in two-electron integral
calculations, as a result of Coulomb operator resolutions. This is
particularly significant in that it eliminates the arduous orbital
translations which were necessary until now for exponential type
orbitals. The bottleneck has been eliminated from evaluation of
three- and four- center integrals over Slater type orbitals and
related basis functions.

The two-center integrals are replaced by sums of overlap-like
one-electron integrals. This implies a speed-up for all basis sets,
including gaussians. The improvement is most spectacular for
exponential type orbitals. A change of basis set is also facilitated
as only these one-electron integrals need to be changed. The
gaussian and exponential type orbital basis sets are, therefore
interchangeable in a given program. The timings of exponential type
orbital calculations are no longer significantly greater than for a
gaussian basis, when a given accuracy is sought for molecular
electronic properties.

Atomic orbitals are physically meaningful one-electron atom
eigenfunctions for the Schr\"odinger equation. This gives well-known
analytical expressions: hydrogen-like orbitals.

Boundary conditions allow the principal quantum number n to be
identified as the order of the polynomial factor in the radial
variable. It must therefore be positive and finite. It is also
defined such that $n \; -l \; -l$ is greater than or equal to 0.
This gives the number of zeros of the polynomial (radial nodes).
Here, $l = 0$ or a positive integer, which defines the angular
factor of the orbital. (i.e. a spherical harmonic, or, more rarely,
its Cartesian equivalent) The number $n$ gives the energy of the
one-electron atomic bound states. Frequently, basis set studies
focus on the radial factor. That is, for our present purposes, the
angular factor can be assumed sufficiently defined as a spherical
harmonic.

The key issue is whether to choose basis sets with exponential or
gaussian asymptotic factors.

Certain physical properties, such as NMR shielding tensor
calculations directly involve the nuclear cusp and correct treatment
of radial nodes, which indicates that basis sets such as Coulomb
Sturmians are better suited to their evaluation than
gaussians~\cite{Coh,Hog2,Lilian}.

There is also evidence to suggest that CI expansions converge in
smaller exponential basis sets compared to
gaussians~\cite{SMILES2,Molp}. Benchmark overlap similarity work is
available~\cite{RAM,LHog}

\section{Wave-function quality}

The following quantity: $$ - 1/2 { \nabla {\rho(r)} \over
{\rho(r)}}$$ is used to test wave-function quality. It is smooth, to
varying degrees, in different basis sets. Atomic positions must give
cusps. The importance for stable and accurate kinetic energy terms, particularly in DFT.

Much molecular quantum chemistry is carried out using gaussian basis
sets and they are indeed convenient and lead to rapid calculations.
The essential advantage they had over exponential basis sets was the
simple product theorem for gaussians on two different atomic
centers. This allows all the two-electron integrals, including
three- and four- center terms to be expressed as single-center
two-electron integrals. \vskip4mm The corresponding relationship for
exponential type orbitals generally led to infinite sums and the
time required, particularly for four-center integrals could often
become prohibitive.

Recent work by Gill has, nevertheless been used to speed up all
three and four center integral evaluation, regardless of basis using
the resolution of the Coulomb operator~\cite{Gill1,Gill2,Hog1}. This
work by Gill is used here to reduce the three- and four- center
two-electron integrals to a sum of products of overlap-like
(one-electron) integrals, basically two-centered. This algorithm was
coded in a Slater type orbital (STO) basis within the framework of
the STOP package~\cite{bouf0} (in fortran) during summer 2008. Note,
however, that other exponential or gaussian basis sets can readily
be used. The set of one-electron overlap-like auxiliary integrals is
the only calculation that needs to be re-done. They may be
re-evaluated for the basis set that the user selects for a given
application. This procedure makes the approach highly versatile,
since a change of basis set requires relatively few simple new
evaluations. A modular or object oriented program is being designed
to do this efficiently~\cite{Hog1,Pin1,Pin2}.

The present article gives illustrative test results on molecular
systems e.g. the H$_2$ dimer. \vskip4mm
 The layout is as follows:
the review begins with a brief recap of basis sets and programming
strategy in the next two sections. Atom pairs are the physical
entity used for integral evaluation, both in the Poisson equation
technique and the Coulomb resolution. Two sections are devoted to
these progressively more powerful techniques which both reduce
two-electron to one-electron integrals. The overlaps required for
the Coulomb resolution differ by a potential factor from orbital
overlaps. Their evaluation is nevertheless analytic, using
well-known techniques summarized in the subsequent section. Finally,
to illustrate what can be gained by eliminating orbital
translations, the translation of Slater type orbitals is reviewed
briefly, from recent work on BCLFs.  A few
numerical results are given on the dimer of molecular hydrogen which
show progressive speed-up particularly for the Coulomb resolution
given a pre-selected accuracy, which proves sufficient to provide
satisfactory confirmation of experimental work on this dimer.
\vskip4mm

\section{Basis sets}
\vskip2mm

Although the majority of electronic quantum chemistry uses gaussian
expansions of atomic orbitals~\cite{Boy,BShav}, the present work
uses exponential type orbital (ETO) basis sets which satisfy Kato's
conditions for atomic orbitals: they possess a cusp at the nucleus
and decay exponentially at long distances from
it~\cite{Kato1,Kato2,Kato3}. It updates a tradition beginning around 1970
and detailed
elsewhere~\cite{Eolo,Stev,ADF0,Shavit,Wils,Clem,Sutt,Poly,Prit,G70,ALCH,Popb,HJ}

Two types of ETO are considered here: Slater type orbitals (STOs)~\cite{STO1,STO2}
and Coulomb Sturmians and their generalisation, which may be written as a finite combination
thereof~\cite{Weni}. Otherwise, STOs may be treated as multiple zeta
basis functions in a similar way to the approach used with gaussian
functions. \par
    \par
    Many exponential type functions exist~\cite{Weni}.
    Preferential use of Sturmian and related functions with similar radial nodes is discussed~\cite{Hog1}.

Coulomb Sturmians have the advantage of constituting a complete set
without continuum states because they are eigenfunctions of a
Sturm-Liouville equation involving the nuclear attraction potential
i.e. the differential equation below.
$$
\nabla_{\vec r}^2 {\cal S}_{nl}^{m}(\beta, {\vec r}) = \big[ \beta^2
- {{2 \beta n}\over{r}} \big] {\cal S}_{nl}^{m}(\beta, {\vec r}).
$$

    The exponential factor of Coulomb Sturmians; $e^{-\beta  r}$ has an arbitrary screening parameter $\beta$.
    In the special case when $\beta$ = $\zeta$/n with n the principal quantum number and $\zeta$ the Slater
    exponent,
    we obtain hydrogen-like functions, which do not span the same space and require
    inclusion of continuum states to form a complete set~\cite{Weni}.
    Hydrogen-like functions are, however well known as atomic orbitals: the radial factor contains the
    associated Laguerre polynomial of order $2l+1$ with suffix
    $n-l-1$ and the exponential $e^{-\zeta r/n}$ as indicated above. The
    angular factor is just a spherical harmonic of order $l$. These
    functions are ortho-normal.
    The optimal values of the $\beta$ parameters may be determined analytically by setting up
    secular equations which make use of the fact that the Sturmian eigenfunctions also
    orthogonalise the nuclear attraction potential, as developed by Avery~\cite{avery2007}.

$$
\int \ {\cal S}_{nl}^{m}(r,\theta,\phi) \, {\cal S}_{n'l'}^{m'}(
r,\theta,\phi) \, {{dr} \over {r}} \, = \,{\delta_{nn'll'mm'}}.
$$

These ortho-normal functions are further generalised by varying
$\alpha$ from the Coulomb Sturmian value of 1. In such a case, the
basis remains ortho-normal and othogonalises $a/r^\alpha$. This
eliminates the $r^2$ term, arising for quadrupole moments when
$\alpha = -2$, thus confirming the very recent numerical observations in the Guseinov group~\cite{Sek}. Similarly, it would be expected that $\alpha = -1$ ETOs
would constitute the optimal basis for magnetic dipole integrals of
NMR shielding. Furthermore, a negative value of $\alpha$ will not
modify the number of radial nodes: the functions will simply breath.

Definitions: the generalised exponential functions constitute finite complete orthonormal sets.

Their expression is as follows:

\begin{equation}
\chi _{nlm}(\mathbf{r})= \, {[\frac{(-1)}{\sqrt{2n}}]}^{\alpha} N_{nl}L_{n-l-1- \alpha}^{2l+2- \alpha}\left( 2 \, \zeta r\right)
r^le^{-\zeta r}Y_l^m(\theta ,\phi )
\end{equation}

Here, N is the normalisation constant previously obtained for
Coulomb Sturmians, L is the associated Laguerre polynomial of order
$2l+2 - \alpha$ with suffix
    $n-l-1 -\alpha$ (recall that $\alpha =1$ defines the Coulomb Sturmians.

Define a variable including the screening constant:

$$ x = 2 \, \zeta r $$

Subsequently, rewriting the norm as $N(\alpha)_{nl}$ and introducing $p = 2l+2- \alpha$ and $q =n+l+1- \alpha$
gives the simplified expression for the generalised orthonormal basis sets of ETO, used by Guseinov.

\vskip4mm

\begin{equation}
\chi _{nlm}(\mathbf{x})= \,  N (\alpha)_{nl}L_q^p\left( x \right)
r^le^{-x/2}Y_l^m(\theta ,\phi )
\end{equation}

In past applications, no obvious advantage has been evidenced for the functions with negative $\alpha$ indices over the well-known Slater type orbitals, Coulomb Sturmians ($\alpha = 1$) or Shull-Loewdin functions ($\alpha=0$). In fact, the infinite series arising when Hartree-Fock two-electron integrals that do not possess closed forms (three and four center terms) are evaluated converge much more slowly when the negative $\alpha$ functions are used. This has recently also proven to be the case of a set of electric field integrals \cite{Zaim}.


This paper records the precedent of electric quadrupole integrals, already published by Guseinov's colleagues, where the negative $\alpha$ basis converges as well as (if not better than) the STO \cite{Sek} and presents a new application to the dipole integrals in the NMR experimental setup.

The investigations are extended to comparisons with previous work on the nuclear dipole integrals that are so important to the evaluation of nuclear shielding tensors and NMR chemical shifts. Furthermore, these nuclear magnetic dipole integrals are closely related to the one-electron nuclear attraction integral, required in all Hartree-Fock and DFT work.

    Alternative ETOs would be Slater type orbitals and B-functions with their simple Fourier
    transforms.
    Strictly, they should be combined as linear combinations to form hydrogen-like or, better,
    Sturmian basis sets prior to use.

STOs allow us to use routines from the STOP package~\cite{QCPE}
directly, whereas Coulomb Sturmians still require some coding. The
relationship to STOs is used to carry out calculations over a
Coulomb Sturmian basis with STOP until the complete Sturmian code is
available. The present state-of-the-art algorithms require at most
twice as long long per integral than GTO codes but the CI converges
with fewer functions and the integrals may be evaluated after
gaussian expansion or expressed as overlaps to obtain speed
up~\cite{Over}.
 \vskip4mm

 \vskip4mm After a suitably accurate electron
density has been obtained for the optimized geometry over a Coulomb
Sturmian basis set, the second order perturbation defining the
nuclear shielding tensor should be evaluated in a Coupled perturbed
Hartree Fock scheme.
\vskip4mm The integrals involved may
conveniently be evaluated using B-functions with linear combinations
giving the Coulomb Sturmians. \vskip4mm
$$ {\cal S}_{nl}^m (r) \, = \, (2 \alpha)^{3/2} {{2^{2 l +1}} \over
{{2l+1)!!}}} \sum_{l=0}^{n-l-1} {{(-n+l+1)_t \, (n+l+1)_t} \over {t!
\, (l+3/2)_t}} B_{t+1,l}^m (r)
$$
\vskip4mm The techniques exploit properties of Fourier transforms of
the integrand.

  Note that either HF or DFT can serve as zero order for the present nuclear shielding tensor calculation over ETOs.

  A full {\it ab initio} B-function code including nuclear shielding tensor work is expected to be complete
  shortly.

  Some tests show that Slater type orbitals (STO) or B-functions (BTO) are less
  adequate basis functions that Coulomb Sturmians, because only the
  Sturmians possess the correct nuclear cusp and radial behavior.

\section{Programming strategy}
Firstly, the ideal {\it ab initio} code would rapidly switch from
one type of basis function to another. Secondly, the chemistry of
molecular electronic structure must be used to the very fullest
extent. This implies using atoms in molecules (AIM) and diatomics in
molecules (DIM) at the outset, following Bader (in an implementation
due to Rico et al~\cite{Ric1} and Tully~\cite{Tull} implemented in
our previous work~\cite{QCPE}, respectively. The natural choice of
atomic orbitals, i.e. the Sturmians or hydrogen-like orbitals lend
themselves to the AIM approach. To a good approximation, core
eigenfunctions for the atomic hamiltonian remain unchanged in the
molecule. Otherwise, atom pairs are the natural choice, particularly
if the Coulomb resolution recently advocated by Gill is used. This
leads us to products of auxiliary overlaps which are either
literally one- or two- centered, or have one factor of the product
where a simple potential function needs to be translated to one
atomic center. The Slater basis set nightmare of the Gegenbauer
addition theorem is completely avoided. Naturally, the series of
products required for, say a four-center two-electron integral may
require 10 or even 20 terms to converge to chemical accuracy, when
at least one atom pair is bound but the auxiliaries are easy to
evaluate recursively and re-use. Unbound pairs may be treated using
a smaller number of terms since the integals can be predicted to be
small, using a Schwarz inequality.

Now, the proposed switch in basis set may also be accomplished just
by re-evaluating the auxiliary overlaps. Furthermore, the exchange
integrals are greatly simplified in that the products of overlaps
just involve a two-orbital product instead of a homogeneous density.
The resulting cpu-time growth of the calculation is $n^2$ for SCF,
rather than $n^4$. Further gains may be obtained by extending the
procedure to post-Hartree-Fock techniques involving explicit
correlation, since the ${r_{12}}^{-1}$ integrals involving more than
two electrons, that previously soon led to bottlenecks, are also
just products of overlaps.

\section{Atom pairs: solving Poisson's equation}

\vskip6mm All the molecular integrals over CS required for standard
SCF may be evaluated using analytical two-center terms based on the
solution of Poisson's equation for the Coulomb potential in an ETO
basis. This uses the Spectral forms (involving incomplete gamma
functions and regular and irregular solid harmonics) defined
initially in~\cite{Red,Over,Man} and subsequently generalized to
ensure numerical stability as shown in a brief summary below.
\par
\vskip1mm
 Recalling the definition of a Slater Type Orbital:
\begin{eqnarray*} \chi_{n,\ l,\ m,\ \zeta}(r,\theta,\phi) = N_1 \; r^{n-1}
e^{-\zeta r} \; Y_l^m(\theta,\phi) \;.
\end{eqnarray*}
\par
Define the radial factor $g(r)$ :
\begin{eqnarray*} g(r) =  r^{n-1}
e^{-\zeta r} \;.
\end{eqnarray*}
\par
Then, (from the Spectral forms in~\cite{Over}), the potential due to
this distribution is immediately written:
$$
\Pi_l(g) =  r^2 F(r) \; ,
$$
\par
Where g is short for g(r) and F(r) is given below, with a suitable
variable of integration; u:
\par
$$
 F(r) = {\int_0^1 du \; g(ru) \ u^{l+2}} +
{\int_1^{\infty} du \; g(ru)\ u^{1-l}.}
$$
\par
This expression is used to write all radially dependent one and
two-center integrals in analytical closed form.

The next section describes a more profound advance, that reduces the
atom-pair evaluation to one-electron overlap-like integrals. It is
related to the Poisson equation technique, as detailed
in~\cite{Over} and~\cite{RinH}.

\section{Avoiding ETO translations for two-electron integrals over 3 and 4
centers}

Previous work on separation of integration variables is difficult to
apply, in contrast to the case for gaussians~\cite{Cesc}
cf~\cite{MH-G}. Recent work by Gill {\it{et al}}~\cite{Gill1}
proposes a resolution of the Coulomb operator, in terms of potential
functions $\phi_i$, which are characterized by examining Poisson's
equation. In addition, they must ensure rapid convergence of the
implied sum in the resulting expression for Coulomb integrals
$J_{12}$ as products of "auxiliaries" i.e. overlap integrals, as
detailed in~\cite{Gill1}.

 This is based on separating the variables of ${1 \over r_{12} }$, by determining suitable functions of $r_1$ and $r_2$ that treat these variables equivalently and constitute a complete set which orthogonalises the Coulomb operator. The associated potentials $\phi_i$ provide an expansion, or resolution of ${1 \over r_{12} }$ similar to that of the identity (using the summation convention):
\begin{equation} \label{lapfn}
| \, g_i \, > \, < \, g_i \,| = I
\end{equation}
This is the completeness property of a set of orthonormal functions, within a particular Hilbert
space. Similarly, for the Coulomb operator, suitable potentials give:
\begin{equation} \label{lapfn}
{1 \over r_{12}} =  | \, \phi_i \, > \, < \, \phi_i \,|
\end{equation}
This Coulomb resolution is based on a complete set of functions which may be determined such that they impose the identity as matrix representation of the Coulomb operator, ${1 \over r_{12} }$ in this basis:

\begin{equation} \label{lapfn}
 < \, f_i \,| {1 \over r_{12} }| \,f_j \,> \, = \delta_{ij}
\end{equation}
The completeness relation for the associated potentials can also be written in the form (8.19)
The functional expression of the above gives:
\begin{equation} \label{lapfn}
{1 \over r_{12}}= \phi_i (r_1) \, \phi_i (r_2) \,
\end{equation}

The potential
functions $\phi_i$, are solutions of Poisson's
equation.
The functions chosen may also be based on Coulomb Sturmians (see the work by Avery).





Completeness of the functions $ f_i$ allows us to expand a density in terms of them:
\begin{equation} \label{lapfn}
< \rho (r) \; | = < \rho (r) \; | {1 \over r_{12} }|
f_i (r)> \; < f_i (r) |
\end{equation}
J is re-written:
\begin{equation} \label{lapfn}
J_{12} \; = \; < \rho (r_1) \; | {1 \over r_{12} }| \rho(r_2) > = \; < \rho (r_1) \; | {1 \over r_{12} }|
f_i (r_1)> \; < \, f_i (r_1) \,| {1 \over r_{12} }| \,f_j (r_2) \,> \,< f_j (r_2) | {1 \over r_{12} }|
\;\rho (r_2) >
 \;
\end{equation}
summing over i and j.
Introducing the orthogonalised operator from 8.20
to resolve the two-
electron integral into a sum of products of 1-electron overlap-like integrals:
\begin{equation} \label{lapfn}
J_{12} \; = \; < \rho (r_1) \; | {1 \over r_{12} }|
f_i (r_1)> \; < f_i (r_2) | {1 \over r_{12} }|
\;\rho (r_2) >
 \; \mbox{with implied summation over $i$}
\end{equation}
And recalling the defining relation for potentials (i.e. one electron functions of a single radial variable):
\begin{equation} \label{lapfn}
| {1 \over r_{12} }| f_i (r)>  = | \phi_i (r) >
\end{equation}

\begin{equation} \label{lapfn}
J_{12} \; = \; < \rho (r_1) \; \phi_i (r_1)> \; < \phi_i (r_2)
\;\rho (r_2) >
 \; \mbox{with implied sumation over i}
\end{equation}
This technique can be readily generalized to exchange and
multi-center two-electron integrals.

Note, however, that the origin of one of the potential functions
only may be chosen to coincide with an atomic (nuclear) position.

\vskip4mm

Define potential functions $\phi_i$ in the scope of a Coulomb
operator resolution, as follows:
\begin{equation} \label{lapfn}
\phi_{n\,l} (r) = \int_0^{+\infty} h_n (x) j_l (rx) dx \;\; \mbox{
with } \; j_l (x) \; \; \mbox{denoting the spherical Bessel
function}
\end{equation}
Here, $h_n (x)$ is the n$^{th}$ member of any set of functions that
are complete and orthonormal on the interval $[0,+\infty)$, such as
the n$^{th}$ order polynomial function (i.e. polynomial factor of an
exponential). The choice made in~\cite{Gill1} is to use parabolic
cylinder functions (see also another application~\cite{Bgauss}),
i.e. functions with the even order Hermite polynomials as a factor.
This is not the only possibility and a more natural and convenient
choice is based on the Laguerre polynomials $ L_n(x)$: Define:
\begin{equation} \label{lapfn}
h_n (x) = \sqrt{2} \; L_n(2 \; x) e^{-x}
\end{equation}
These polynomial functions are easy to use and lead to the following
analytical expressions for the first two terms in the potential
defined in (6.2):
\begin{equation} \label{lapfn}
V_{00} (r) = \sqrt{2} \; {{ \mbox{tan}^{-1} (r)} \over {r}}
\end{equation}

\begin{equation} \label{lapfn}
V_{10} (r) = \sqrt{2} \; [ \; {{ \mbox{tan}^{-1} (r)} \over {r}} \;-
\; { 2 \over {(1 \; + r^2)}}]
\end{equation}
\vskip4mm

Furthermore, higher $n$ expressions of $V_{n 0}(r)$ all resemble
(6.5) (see~\cite{Gill2} eq (23)):
\begin{equation} \label{lapfn}
V_{n 0}(r) = \sqrt{2} \; { 1 \over {r}} (1 + \sum_1^n (-1)^k
{{\mbox{sin} (2 \; k \; \mbox{tan}^{-1} (r))} \over k})
\end{equation}
and analytical expressions of $V_{n l}(r)$ with non-zero $l$ are
also readily obtained by recurrence.

The auxiliary overlap integrals $< \rho (r_1) \; \phi_i (r_1)>$ and
$< \phi_i (r_2) \; \rho (r_2) >$ will involve densities obtained
from atomic orbitals centered on two different atoms in most
multi-center two-electron integrals. The integrals required in an
ETO basis are thus of the type:
\begin{equation} \label{lapfn}
< \psi_a (r_1) \; \psi_b (r_1) \; \phi_i (r_1) >
\end{equation}

Such integrals appear for two-center exchange integrals and all
three- and four center integrals. Note that exchange integrals
require distinct orbitals $\psi_a$ and $\psi_b$. In the atomic case,
they must have different values for at least one of $n,l,m \;
\mbox{or}\;  \zeta$. In the two-center case, the functions centered
at a and b may be the same. The product does not correspond to a
single-center density: it is two-centered. The above equation then
illustrates the relationship to the one-electron two-center overlap
integral, although it clearly includes the extra potential term from
the Coulomb operator resolution.

\vskip4mm The overlap integrals may be evaluated by separating the
variables in prolate spheroidal co-ordinates, following Mulliken and
Roothaan~\cite{Roo} and using recurrence relations in~\cite{HTE}:
\vskip8mm
$$ S(n \sb{1},l \sb{1},m,n \sb{2},l \sb{2},\alpha ,\beta )=\alpha \sp{n
\sb{1}+1/2}\beta \sp{n \sb{2}+1/2} \left\lbrack \left(2n \sb{1}
\right)! \left(2n \sb{2} \right)! \right\rbrack \sp{-1/2}s(n \sb{1}l
\sb{1}mn \sb{2}l \sb{2}\alpha \beta ) \qquad $$ \vskip8mm
$$ =N(n \sb{1},n \sb{2},\alpha \beta )s(n \sb{1},l \sb{1},m,n \sb{2},l
\sb{2},\alpha \beta ) $$ \vskip8mm \noindent where: $ \alpha =k
\sb{1}R $ and $ \beta =k \sb{2}R $. The $ k \sb{1},k \sb{2} $ are
Slater exponents. \vskip8mm The core overlaps are given by:
\vskip2mm
$$ s(n \sb{1},l \sb{1},m,n \sb{2},l \sb{2},\alpha ,\beta )= \int \sp{\infty}
\sb{1} \int \sp{1} \sb{-1}exp \left\lbrace -{1 \over 2}(\alpha
+\beta )\mu -{1 \over 2}(\alpha -\beta )\nu \right\rbrace (\mu +\nu
) \sp{n \sb{1}}(\mu -\nu ) \sp{n \sb{2}}  T(\mu ,\nu )d\mu d\nu
$$ \vskip2mm
$$ \mu ={r \sb{a}+r \sb{b} \over R} $$
\vskip2mm
$$ \nu ={r \sb{a}-r \sb{b} \over R} $$
\vskip2mm \indent$ r \sb{a} $ and $ r \sb{b} $ are the instantaneous
position vectors of the electron from the two centers labeled a and
b, respectively and separated by a distance R. We also define, suing
the normalised spherical tensors S: \vskip8mm
$$ T(\mu ,\nu )=S \sp{m} \sb{l \sb{1}}(\mu ,\nu ) \sb{a}S \sp{m} \sb{l
\sb{2}}(\mu ,\nu ) \sb{b} $$.
\par \vskip4mm \indent The core overlaps then take the form: \vskip8mm
$$ s(n \sb{1},l \sb{1},m,n \sb{2},l \sb{2},\alpha ,\beta )=D \sb{l \sb{1},l
\sb{2},m} \sum \sp{\lambda} \sb{ij}Y \sp{\lambda} \sb{ij}A \sb{i}
\left\lbrace{ 1 \over 2}(\alpha +\beta ) \right\rbrace B \sb{j}
\left\lbrace{ 1 \over 2}(\alpha -\beta ) \right\rbrace \quad $$
\vskip8mm \noindent$ Y \sp{\lambda} \sb{ij} $ is a matrix with
integer elements uniquely determined from $ n, \; l \; \mbox{and} \;
m$. It is obtained as a generalised binomial coefficient, in the
expansion of $(r_a-r_b)^n \, (r_a+r_b)^n$\vskip4mm $ D \sb{l
\sb{1},l \sb{2},m} $ \hskip2mm is a coefficient that is independent
of the principal quantum number. It is obtained upon expanding the
product of two Legendre functions in this co-ordinate system.
Symmetry conditions imply that only $m_1 \, = \, m_2 \, = \, m$ lead
to non-zero coefficients. \vskip8mm
$$ A \sb{i} \left\lbrace{ 1 \over 2}(\alpha +\beta ) \right\rbrace = \int
\sp{\infty} \sb{1}exp \left\lbrace -{1 \over 2}(\alpha +\beta )\mu
\right\rbrace \mu \sp{i}d\mu $$ \vskip12mm
$$ B \sb{j} \left\lbrace{ 1 \over 2}(\alpha -\beta ) \right\rbrace = \int
\sp{1} \sb{-1}exp \left\lbrace -{1 \over 2}(\alpha -\beta )\nu
\right\rbrace \nu \sp{j}d\nu $$ \vskip12mm

Here, recurrence relations on the auxiliary integrals A and B lead
to those for the requisite core integrals~\cite{HTE, Gus1}.

 This assumes tacitly that the potential obtained from the coulomb operator resolution be
centered on one of the atoms. Whilst this choice can be made for one
pair in a four-center product, it cannot for the second. There
remains a single translation for this potential in one auxiliary of
the two in a product representing a four-center integral and none
otherwise. The structure of these potential functions in (6.5) and
(6.6) shows that the translation may be accomplished readily in the
prolate spheroidal co-ordinates. This point is addressed in detail
in a submitted manuscript~\cite{Hog1}.

This method obviates the need to evaluate infinite series that arise
from the orbital translations efficiently. They have been eliminated
in the Coulomb operator resolution approach, since only orbitals on
two centers remain in the one-electron overlap-like auxiliaries.
These can be evaluated with no orbital translation, in prolate
spheroidal co-ordinates, or by Fourier
transformation~\cite{Hog1,Gill2}. \vskip5mm
\section{Numerical results compared for efficiency}
Consider the H$_2$ molecule and its dimer/agregates. In an s-orbital
basis, all two-center integrals are known analytically, because they
can be integrated by separating the variables in prolate spheroidal
co-ordinates. A modest s-orbital basis is therefore chosen, simply
for the demonstration  on a rapid calculation, for which some
experimental data could be corroborated. \par The purpose of this
section is to compare evaluations using the translation of a Slater
type orbital basis to a single center (STOP)~\cite{QCPE} with the
Poisson equation solution using a DIM (Diatomics in molecules or
atom pair) strategy and finally to show that the overlap auxiliary
method is by far the fastest approach, for a given accuracy (the
choice adopted is just six decimals, for reasons explained below).
\par H$_2$ molecule with interatomic
distance of 1.402d0 atomic units (a.u.) The first table (Table 1)
assembles the full set of all Coulomb integrals; with one and
two-centers evaluated using STOP, Poisson and overlap methods.
Exponents may be found from the atomic integrals which do not
include the constant factor (5/8 here).

\vskip10mm
\newpage
Table 1. \par

\begin{tabular}{|c|c|c|c|c|}
  \hline
  AOs (zeta) & $1s_{a1}$ & $1s_{a2}$ & $2s_{a1}$ & $2s_{a2}$ \\
  \hline
  $1s_{a1}$ 1.042999 & 1.042999 & 0 & 0 & 0 \\
  $1s_{a2}$ 1.599999 & 0.934309 & 1.599999 & 0 & 0 \\
  $2s_{a1}$ 1.615000 & 0.980141 & 0.870304 & 1.615000 & 0 \\
  $2s_{a2}$ 1.784059 & 0.901113 & 0.923064 & 1.189241 & 1.784059 \\
  $1s_{b1}$ 1.042999 & 3.455363 & 0.364117 & 0.659791 & 1.621644 \\
  $1s_{b2}$ 1.599999 & 0.433097 & 0.332887 & 0.635867 & 1.541858 \\
  $2s_{b1}$ 1.615000 & 0.323691 & 0.248050 & 0.529300 & 1.276630 \\
  $2s_{b2}$ 1.784059 & 0.402387 & 0.324872 & 0.636877 & 2.014196 \\
  \hline
\end{tabular}
\vskip10mm Table 2. \par

Atomic exchange integrals (6 distinct single center values between
pairs of different AOs).\par \vskip5mm
\begin{tabular}{|c|c|c|c|c|}
  \hline
  AOs (zeta) &  Label  & [a(1)b(2)a'(2)b'(1)] & value & comment \\
  \hline
  $1s_{a1}$ 1.042999 & 1 & 1212 & 0.720716 & - \\
  $1s_{a2}$ 1.599999 & 2 & 1313 & 0.585172 & - \\
  $2s_{a1}$ 1.615000 & 3 & 1414 & 0.610192 & - \\
  $2s_{a2}$ 1.784059 & 4 & 2323 & 0.557878 & - \\
  $1s_{b1}$ 1.042999 & 5 & 2424 & 0.607927 & - \\
  $1s_{b2}$ 1.599999 & 6 & 3434 & 0.602141 & - \\
  $2s_{b1}$ 1.615000 & 7 & 2121 & 0.720716 & = 1212 \\
  $2s_{b2}$ 1.784059 & 8 & 3232 & 0.585172 & = 2323 \\
  \hline
\end{tabular}
\par
\vskip5mm Table 3.
\par
Two-center exchange integrals. All pair permutations possible. Some
are identical by symmetry. \vskip5mm
\begin{tabular}{|c|c|c|}
\hline Labels & value & comment \\

\hline

1515 & 0.319902 & - \\
1516 & 0.285009 & = 1525 \\
1517 & 0.325644 & = 1535 \\
1518 & 0.324917 & = 1545 \\
1527 & 0.291743 & = 1536 \\
1528 & 0.293736 & = 1547 \\
1538 & 0.329543 & = 1548 \\
2525 & 0.260034 & - \\
2516 & 0.254814 & - \\
2517 & 0.290533 & - \\
2518 & 0.290149 & - \\
\hline
\end{tabular}

 \vskip4mm
The two-center integrals are dominated by an exponential of the
interatomic distance and thus all ave values close to 0.3. The table
is not the full set. All `15` terms, involving $1s_{a1}(1) \;
1s_{b1}(2)$ are given, to illustrate symmetry relations.

Note that this is by no means the best possible basis set for H$_2$,
since it is limited to $l=0$ functions (simply to ensure that even
the two-center exchange integral has an analytic closed form).
\par
\newpage

The total energy obtained for the isolated H$_2$ molecule is
-1.1284436 Ha as compared to a Hartree-Fock limit estimate of
-1.1336296 Ha. Nevertheless, the Van der Waals well, observed at 6.4
au with a depth of 0.057 kcal/mol is quite reasonably
reproduced~\cite{Hdim}.
 \vskip8mm Dimer geometry: rectangular and planar. Distance
between two hydrogen atoms of neighboring molecules: 6 au. Largest
two-center integral between molecules: 4.162864 10$^{-5}$. (Note
that this alone justifies the expression dimer-the geometry
corresponds to two almost completely separate molecules, however,
the method is applicable in any geometry).

Timings on a Power 6 workstation, for the dimer (all 4-center
integrals in msec): \par STOP: 12 POISON: 10 OVERLAP: 2. \vskip8mm
Total dimer energy: -2.256998 Ha. This corresponds to a well-depth
of 0.069kcal/mol, which may be considered reasonable in view of the
basis set. The factor limiting precision in this study is the
accuracy of input. The values of Slater exponents and geometric
parameters are required to at least the accuracy demanded of the
integrals and the fundamental constants are needed to greater
precision.

\subsection{The NMR nuclear shielding tensor.}

More complete work is referred to here and the present
description is a brief summary~\cite{Ditchfield1,Ditchfield2,Nabil}.
In NMR, the nuclear shielding tensor is a second order perturbation
energy correction, for derivatives with respect to the nuclear
dipole moment and the external field.

The perturbed Fock matrix element when including the effect of the
external field contains both one and two electron terms. In this
example, we focus on the one electron terms.

The purpose of the present section is to give a case study of one of
the contributing energy integrals involving the dipole $1/r_N^3$
operator.

In the applied magnetic field, the question of gauge invariance must
be resolved. A method of circumventing the problem was devised by
Ditchfield using the London GIAO~\cite{London}. These Gauge
Including Atomic Orbitals reduce to STO for zero field and contain
the required phase factor
otherwise~\cite{Ditchfield1,Ditchfield2,Berlu1}.

The integrals were evaluated for GTO at zero field and nuclear
shielding tensor or chemical shifts have been available since
Gaussian 72 based on this pioneering work~\cite{G70} and distributed
to academics by QCPE. It is nevertheless important that users input
the appropriate structure in order to obtain accurate chemical
shifts corresponding to the species studied and note that for work
in solution (or in solids) some structural changes may occur.

Define the nuclear shielding tensor as a second order energy
perturbation:
\begin{equation}
  \label{eq:3sec1}
  \sigma_{\alpha\beta}^N
  = \left[
    \frac{\partial^2
      \left\langle 0 \left|
          \mathcal{H}(\vec\mu_N, \vec B_0)
        \right| 0 \right\rangle}{\partial\mu_{N, \, \alpha}\,\partial B_{0, \ \beta}}
  \right]_{\vec\mu_N = \vec 0, \ \vec B_0 = \vec 0}
\end{equation}
with $\vec{\mu}_N$ the nuclear dipole moment of nucleus $N$ and
$\vec{B}_0$ the external field. $|0\rangle$ is a closed shell ground
state Slater determinant. $\alpha$ and $\beta$ stand for cartesian
coordinates.

A coupled Hartree-Fock treatment of the above equation leads
to~\cite{Ditchfield1,Ditchfield2,Stevens2}:
\begin{equation}
  \label{eq:16sec1}
  \sigma_{\alpha \beta}^N
  = Tr \left[ P^{(0,1)}_{\beta} h^{(1,0)}_{\alpha} + P^{(0)} h^{(1,1)}_{\alpha \beta} \right]
\end{equation}
where $P^{(0)}$ and $P^{(0,1)}_{\beta}$ are the density matrix of
zero order and first order with respect to the external magnetic
field, $h^{(1,0)}_{\alpha}$ is the core Hamiltonian of the first
order with respect to nuclear dipole moment and $h^{(1,1)}_{\alpha
\beta}$ is the second order one-electron Hamiltonian with respect to
the nuclear moment $\mu_\alpha$ and the external field $B_\beta$.

The non-zero orders in~(\ref{eq:16sec1}) involve integrals which are
absent from {\it ab initio} Hartree-Fock calculations. In this work,
we focus our attention on integrals involving $1/r_N^3$ in their
operator. These integrals appearing in the second order expression
for the approximate perturbed Hamiltonian:
\begin{equation}
  \label{eq:83chap1}
  \begin{split}
    h_{\mu\nu,\, \alpha\beta}^{(1,1)} = \frac{\mu_0}{4\pi}\frac{e^2}{2 m_e}
    \left\langle \chi_\mu \left|
        \frac{\vec{r}_{\nu} \cdot \vec{r}_N \delta_{\alpha\beta} - r_{\nu,\alpha}r_{N,\beta}}{r_N^3}
        +
        \frac{(\vec{R}_{\mu\nu} \wedge \vec{r})_\beta (\vec{r}_{N} \wedge \vec{\nabla})_{\alpha}}{r_N^3}
      \right| \chi_\nu \right\rangle
  \end{split}
\end{equation}
                                %
The integral which we have chosen to investigate in detail within
the Fourier transform approach, is the three-center one electron
integral:
\begin{equation}
  \label{eq:int1}
  I = \left\langle \chi_\mu \left|
      \frac{\vec{r}_{\nu} \cdot \vec{r}_N \delta_{\alpha\beta} - r_{\nu,\alpha}r_{N,\beta}}{r_N^3}
    \right| \chi_\nu \right\rangle
\end{equation}
here $\vec{r}_N$ is the instantaneous position of the electron with
respect to the nuclei $N$. \vskip0.5mm
\centerline{\bf Analytical treatment.} This is to be found in the
Appendix.

The above algorithm is available in Fortran, within the STOP (Slater
Type Orbital Package) set of programs, at the coupled perturbed
Hartree-Fock level with the ETOs axpanded in Slater type orbitals.

DFT coding proves more accurate for NMR chemical shifts because it
accounts for the majority of the electron correlation energy. In
this case, the ETOs are fitted to large Gaussian expansions,
following the algorithm in ~\cite{Bgauss} and Gaussian03 is
subsequently used.\vskip2mm


\centerline{\bf Application.}
\par
\vskip1mm The $^{15}N$ chemical shifts measured for a set of
benzothiazoles are evaluated with the above expressions. These molecules possess a ring nitrogen and have
been studied previously in our group~\cite{Berlu1}. The measurements
were made in natural abundance. The intensity of signals due to the
nitrogen must be amplified by a 2-D NMR technique involving
cross-polarization to benefit from the intensity of proton
resonances coupled to that of the $^{15}N$ in the molecule. \par

The {\it in vivo} NMR benefited from measurements by Bruno
Combourieu: these molecules are metabolized by bacteria and
researchers in the group try to follow the pathway by NMR. Since
such studies are very difficult to do, we tried calculating some
chemical shifts accurately from structures to assign them (see
acknowledgements). \vskip1mm

 The Y substituent, generally a hydroxide was found to be
in the position indicated (for mechanistic reasons, it is the only
accessible and stabilised position for ring hydroxylation which has
been found to take place {\it in vivo} after experiments in our
group).
\vskip1mm

 In solution, these molecules undergo a tautomeric
equilibrium reaction transferring a proton towards this nitrogen as
shown in the figure below (also used for nomenclature; P= protonated
on resonating nitrogen).
\par
\begin{figure}[bhtp]
\begin{center}
\includegraphics{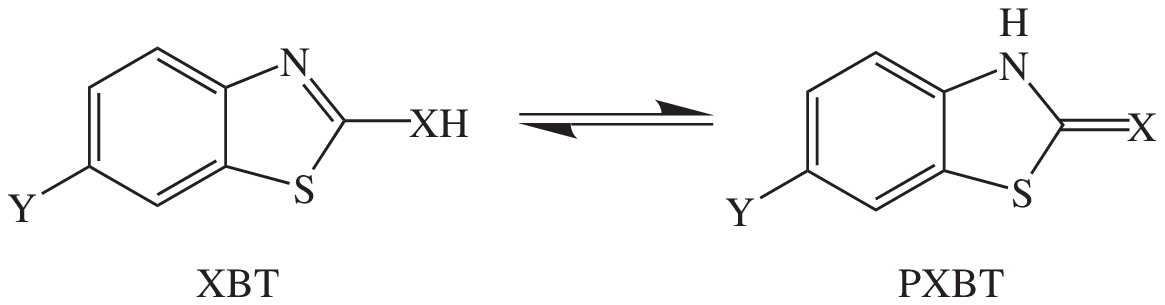}
\end{center}
\label{Benzotaut}
\end{figure}

Summary of NDDO-PM3 fitted STO molecular-site calculations on
unprotonated tautomers (b). It is possible to conclude when the
Gaussian03/PBE 6-311++G(2d,p) calculation (c) differs substantially
from the measured value (a) (ppm/$CH_3NO_2$) that the resonating
nitrogen is mostly protonated. This serves as a guideline for {\it
ab initio} structures studied for these equilibria.
\par
\vskip1mm

 { \begin{tabular}{|c|c|c|c|c|}
  \hline
 molecule&substituent& a&    b &      c \\
\hline
 BT:benzothiazole:&No X&    -72.5   & -71.8   & -61.4  \\
\hline

 OBT:&X=O&-238.8  & -238.9 & -133.3  \\
 \hline
 OHOBT:&X=O; Y=OH &  -240.4 & -239.9 & -135.3 \\
 \hline
ABT:&X=NH&   -153.1 & -152.1 & -131.6  \\
\hline
 OHABT:&X=NH; Y=OH&  -153.1 & -153.6 & -132.3 \\
 \hline
 MBT:&X=S &-199.6 & -199.9   & -79.9 \\
 \hline
  OHMBT:&X=S; Y=OH &-205.5  & -205.5 & -83.2 \\
  \hline
MTB:&X=N(CH$_3$)CONHCH$_3$&-124.0 & -125.4 &-141.0 \\
\hline
\end{tabular}}
\vskip1mm The above results prompted use of a structure, protonated
on the resonating N, (denoted P) to obtain the zero-order
wave-function, in all cases apart from benzothiazole (BT) and ABT.
Below, the same cases are treated in the DFT work.

Note that the basis sets including hydrogen-like orbitals perform
better than the STO basis sets that in turn improve upon dense-core
Gaussian basis sets [6-311++G(2d,p)].

Next, examining the generalised basis sets, compare b,c, d, and f
with the measured chemical shifts and evaluate the difference in
ppm. Note that
\newpage
\vskip4mm {\underline{ TABLE: DFT Calculations.}} Differences
between calculated and observed $^{15}N$ chemical shifts for
commercial benzothiazoles and some metabolites (in ppm).\par

a-Measured values with respect to nitromethane standard in
deuterated methanol solvent \par (B. Combourieu in~\cite{Berlu1})
error bars of 2 ppm.
\par
b-Coupled perturbed STO.
\par c-Gaussian~\cite{G03} augmented with
hydrogen-like AOs (c.f. Coulomb Sturmians $\alpha= 1$.
\par d-Gaussian~\cite{G03}. \par

\par e-Generalised ETO $\alpha= -1$

\par f-Generalised ETO $\alpha= -2$

Note. b through f involve solvation models, detailed below.

\vskip4mm {\begin{tabular}{|c|c|c|c|c|c|c|}
  \hline
 molecule&substituent& a-c &    a-b &       a-d &       a-e &       a-f \\
\hline
 BT:benzothiazole:&No X&       1.3 &      8.3 &     11.1 &     1.2 &     2.8  \\
\hline

 POBT:&X=O&4.6 & 11.7 &     20.0  & 3.8 &  5.3\\
 \hline
 POHOBT:&X=O; Y=OH &                  4.5& 7.4 &14.9& 2.9 & 5.2\\
 \hline
ABT:&X=NH&                 1.1 &3.8&       21.5& 0.9 & 3.1 \\
\hline
 OHABT:&X=NH; Y=OH&4.5 &10.1& 20.8& 2.8 & 6.1\\
 \hline
 PMBT:&X=S &3.0 &11.2 &21.2&2.1 & 4.5\\
 \hline
  POHMBT:&X=S; Y=OH &2.5 &10.1& 18.8&1.7 & 4.8\\
  \hline

\end{tabular}}
\par
\vskip4mm Basis sets augmented with hydrogen-like orbitals are
within 5 ppm of the experimental values (measured within 2 ppm) for
the discrete solvated model. This model explicitly includes several
deuterated methanol molecules to cater for the specific hydrogen
bonding interactions. \vskip4mm
\par a Measured chemical shift for ring nitrogen.
\par b STO: DFT PBE 6-311++G(2d,p) calculations with two
discrete $CD_3OD$ molecules on OH, NH, and SH (one on N, O, S) for
minimal total energy.
\par c Gaussian 2003 as (b) with hydrogen-like orbital DFT PBE
aug-6.311G**(2d, p) calculations. \par d Gaussian 2003 DFT PBE
6-311++G(2d,p) calculations. \par The content of this table is
original and based on the previous work of the author~\cite{Hog1}
i.e. geometries are re-optimized from the co-ordinates
of~\cite{Hog1}.

\vskip10mm

\centerline{\bf Conclusion.}

\vskip4mm

Another step on the way to {\it ab initio} ETO basis nuclear
screening tensor calculations has been accomplished. \par It is
essential to use a basis set which comprises orbitals with the
correct nuclear cusp behavior. This implies a non-zero value of the
function at the origin for spherically symmetric cases and
satisfying Kato's conditions. Hydrogen-like atomic orbital basis
sets therefore perform better than Slater type orbitals which are an
improvement upon even large Gaussian basis sets. \par The NDDO-PM3
molecular site approach has the advantage of rapidity. Calculations
take about a minute instead of 50-75 hours on the IBM-44P-270. They
cannot be systematically improved, however once the site Slater
exponents have been fitted. Note that the 2s Slater exponent
fluctuates wildly in fits, providing further evidence that shielding
must be of the form (2-r) for the 2s ETO.

Fundamental work on orbital translation is also in progress to speed
up these calculations within the test-bed of the STOP
programmes~\cite{CAW,Sidi,Hogab}.

The interplay of these discrete molecule solvent models and accurate
{\it in vivo} NMR measurements is satisfactory, in that the
structures postulated give calculated chemical shifts to similar
accuracy as obtained for experimental values (on the order 2ppm). It
should be stressed that energy minimization in this case does
evidence directional hydrogen bonds but can lead to several possible
solvent geometries. Further study, using molecular dynamics
techniques would be useful in the modeling of solvent shells and is
planned in the future. In view of the complex systems studied, this
is highly satisfactory.

\vskip8mm
\section{Conclusions}
It is a remarkable gain in simplicity that the Coulomb operator
resolution~\cite{Gill1} now enables the exponential type orbital
translations to be completely avoided, although some mathematical
structure has been emerging in the BCLFs used to translate Slater
type orbitals and even more in the Shibuya-Wulfman matrix used to
translate Coulomb Sturmians.

\vskip4mm This breakthrough in the ETO algorithm strategy represents
a well-controlled approximation, analogous to resolutions of the
identity. The convergence has been shown to be rapid in all
cases~\cite{Gill2}.
\section{Acknowledgements}
The author would like to thank Peter Gill for helpful discussions at
ISTCP-VI and rapidly transmitting the text~\cite{Gill2}. 

Didier Pinchon, as usual, contributed to the scientific
debate surrounding this work.
\section{Appendix: analysis of nuclear dipole integrals for NMR in a
Slater basis}

The operator, $\left(\vec{r}_{\nu} \cdot \vec{r}_N
\delta_{\alpha\beta} -
r_{\nu,\alpha}r_{N,\beta}\right)/\left(r_N^3\right)$, can be
expressed as a combination of terms involving cartesian
co-ordinates. These terms take the following general form:
\begin{equation}
  \label{eq:int2}
  \frac{X^{\nu,\, i}X^{N,\, j}}{r_N^3}
\end{equation}
where $X^{N, \, j}$ stands for cartesian coordinates of the electron
with respect to the nuclei $N$.

Now, it is more convenient to express the cartesian co-ordinates as
sums of spherical polar co-ordinates with their complex conjugates.
These co-ordinates are of the general form:
\begin{equation}
  \label{eq:int3}
  \begin{aligned}
    & x = r \sqrt{\frac{2\pi}{3}}
    \bigl[ Y_1^{-1} (\theta_{\vec{r}},\varphi_{\vec{r}}) - Y_1^1 (\theta_{\vec{r}},\varphi_{\vec{r}}) \bigr] \\
    & y = i r \sqrt{\frac{2\pi}{3}}
    \bigl[ Y_1^{-1} (\theta_{\vec{r}},\varphi_{\vec{r}}) + Y_1^1 (\theta_{\vec{r}},\varphi_{\vec{r}}) \bigr] \\
    & z = r \frac{4\pi}{3} Y_1^0(\theta_{\vec{r}},\varphi_{\vec{r}})
  \end{aligned}
\end{equation}

The product of a STO by a cartesian coordinate can be expressed as a
combination of STO, since:
\begin{equation}
  \label{eq:int4}
 \begin{split}
    r Y_L^M(\theta_{\vec{r}},\varphi_{\vec{r}})
    & \chi_{n,l}^{m}(\zeta, \vec{r})
   = r Y_L^M(\theta_{\vec{r}},\varphi_{\vec{r}}) \times r^{n-1} e^{-\zeta r} Y_l^m(\theta_{\vec{r}},\varphi_{\vec{r}})
    & = (-1)^M \sum_{\lambda=\lambda_{min}, 2}^{\lambda_{max}} \left\langle l m \left| L -M \right| \lambda m+M \right\rangle
    r^{n} e^{-\zeta r} Y_{\lambda}^{m+M}(\theta_{\vec{r}},\varphi_{\vec{r}})
    & = (-1)^M \sum_{\lambda=\lambda_{min}, 2}^{\lambda_{max}} \left\langle l m \left| L -M \right| \lambda m+M \right\rangle
    \chi_{n+1, \lambda}^{m+M}(\zeta, \vec{r})
  \end{split}
\end{equation}
here we have used the Gaunt coefficients~\cite{Gaunt1,Gaunt2} and
the Condon and Shortley phase convention for spherical harmonics
$Y_l^m(\theta_{\vec{r}},\varphi_{\vec{r}})$~\cite{Condon}.

Consequently, the integral~(\ref{eq:int1}) is reduced to a sum of
terms of the form:
\begin{equation}
  \label{eq:int5}
  \left\langle \chi_\mu \left | \frac{Y_1^j(\theta_{\vec{r_N}},\varphi_{\vec{r_N}})}{r_N^2} \right| \chi_\nu \right\rangle
\end{equation}
with just a $1/r^2$ dependence and where $j = -1, 0, 1$.\\

Using the Fourier transform formalism requires the integral
representation of the operator involved in~(\ref{eq:int5}). We
obtain:
\begin{equation}
  \label{eq:int6}
  \overline{\left( \frac{Y_1^j(\theta_{\vec{r_N}},\varphi_{\vec{r_N}})}{r_N^2} \right)} (\vec{p})
  = \frac{2i}{\sqrt{2\pi}} \frac{Y_1^j(\theta_{\vec{p}},\varphi_{\vec{p}})}{p}
\end{equation}

This immediately allows us to write the inverse Fourier transform:
\begin{equation}
  \label{eq:int7}
  \frac{Y_1^j(\theta_{\vec{r_N}},\varphi_{\vec{r_N}})}{r_N^2}
  = - \frac{i}{2\pi^2} \int{\frac{e^{-i\vec{p}\cdot\vec{r_N}} Y_1^j(\theta_{\vec{p}},\varphi_{\vec{p}})}{|\vec{p}|} \,d\vec{p}}
\end{equation}

Now, this places us in a position to write the Fourier integral for
the present term in the NMR nuclear shielding tensor calculation.
After expanding the Slater type orbitals in terms of B-functions and
substituting~(\ref{eq:int7}) in~(\ref{eq:int5}), the present
integral becomes:
\begin{equation}
  \label{eq:int8}
  \begin{split}
    \mathcal{I}
    & = - \frac{i}{2\pi^2} \int \frac{e^{i\vec{p}\cdot\vec{R_N}} Y_1^j(\theta_{\vec{p}},\varphi_{\vec{p}})}{|\vec{p}|} \\
    & \quad \times \langle B_{n_1,l_1}^{m_1}(\zeta_1, \vec{r}) | e^{-i\vec{p}\cdot\vec{r}}
    | B_{n_2,l_2}^{m_2}(\zeta_2, \vec{r}-\vec{R_2})\rangle_{(\vec{r})} \,d\vec{p}
  \end{split}
\end{equation}
whereas the three-center nuclear attraction integral is :
\begin{equation}
  \mathcal{I} = \frac{1}{2\pi^2} \int \frac{e^{i\vec{p}\cdot\vec{R_N}}}{|\vec{p}|^2}
  \times \langle B_{n_1,l_1}^{m_1}(\zeta_1, \vec{r}) | e^{-i\vec{p}\cdot\vec{r}}
  | B_{n_2,l_2}^{m_2}(\zeta_2, \vec{r}-\vec{R_2})\rangle_{(\vec{r})} \,d\vec{p}
\end{equation}
\vskip4mm The three-center dipolar integral (13) appears in a form
closely related to that of the three-center nuclear attraction
integrals required at the HF-SCF level of electronic structure
calculation (and also used in electronic DFT work).

 In both above integrals note the presence of the common
factor in B-function Fourier transform work first studied by the
Steinborn group: eq (5.4 and 5.5). See also \cite{Steinborngroup}
i.e.:
\begin{equation}
  \langle B_{n_1,l_1}^{m_1}(\zeta_1, \vec{r}) | e^{-i\vec{p}\cdot\vec{r}}
  | B_{n_2,l_2}^{m_2}(\zeta_2, \vec{r}-\vec{R_2})\rangle_{(\vec{r})}
\end{equation}
The analytical treatment developed here has not required any
hypothesis on the relative position of nucleus (aligned or not) and
any restriction on quantum numbers. Consequently, the
equation~(\ref{eq:int8}) is completely general and may be directly
evaluated from routines available in a quantum calculation software.

Note that such an integral satisfies all applicability conditions of
non-linear transformations for extrapolation described by A.
Sidi~\cite{Sid}. Previous work on three-center nuclear integral
evaluation~\cite{Safhog} has been used to develop an efficient
program to compute this dipolar $1/r_N^3$ three-center integral.

\end{document}